\documentclass[journal,onecolumn,11pt]{IEEEtran}

\usepackage{amsmath,amssymb}
\usepackage[ps,dvips,all,import,arc]{xy}

\allowdisplaybreaks

\newcommand{\Z}{\mathbb{Z}}

\newcommand{\gd}{\delta}

\renewcommand{\phi}{\varphi}

\newtheorem{thm}{Theorem}
\newtheorem{lem}[thm]{Lemma}

\newcommand{\set}[1]{\{#1\}}

\begin{document}

\title{Distance-Increasing Maps of All Length \\ by Simple Mapping Algorithms}

\author{Kwankyu~Lee,~\IEEEmembership{Member,~IEEE}
\thanks{This work was supported by the Korea Research Foundation Grant funded by Korea Government (MOEHRD, Basic Research Promotion Fund) (KRF-2005-214-C00009).}%
\thanks{The author is with the Department of Mathematics, San Diego State University, San Diego, CA 92182 USA
(e-mail: kwankyu@sogang.ac.kr).}}

\maketitle

\begin{abstract}
Distance-increasing maps from binary vectors to permutations, namely DIMs, are
useful for the construction of permutation arrays. While a simple mapping algorithm defining DIMs of even length is known, existing DIMs of odd length are either recursively constructed by merging shorter DIMs or defined by much complicated mapping algorithms. In this paper, DIMs of all length defined by simple mapping algorithms are presented. 
\end{abstract}

\IEEEpeerreviewmaketitle


\section{Introduction}
Let $S_n$ be the set of all permutations of $\set{1,2,\dots,n}$. Here we think of a permutation $x=(x_1,x_2,\dots,x_n)$ as a tuple rather than a map.  Let $\Z_2^n$ be the set of all binary vectors of length $n$. Both $S_n$ and $\Z_2^n$ are endowed with the Hamming distance $d$.

A distance-increasing map of length $n$, $n$-DIM for short, is a map $f$ from $\Z_2^n$ to $S_n$ that increases the Hamming distance, that is, $d(f(u),f(v))> d(u,v)$ for all $u$, $v\in\Z_2^n$ except when $d(u,v)=n$. DIMs were first studied because of their useful application in constructing permutation arrays. The image of a binary code under a DIM is a permutation array with minimum distance greater than that of the binary code. However, DIMs are also interesting combinatorial objects. It is easy to see by a counting argument that no DIMs of length $<4$ can possibly exist. First known examples of DIMs are $h_{2m}$ in \cite{chang_2003} for $m=2$ and odd $m\ge3$. What is interesting with $h_{2m}$ is that they are defined by a very simple mapping algorithm that returns a permutation for a binary vector given as input. Then Chang \cite{chang:2005} succeeded in constructing DIMs of all length by recursively merging DIMs of shorter length, beginning with a small set of basic DIMs. The basic DIMs included some maps belonging to the family $h_{2m}$ and a computer-found one. Recently a mapping algorithm as simple as that of $h_{2m}$ defining DIMs of all even length was found \cite{kwankyu:2005}. Thus a natural question arose whether there exist DIMs of odd length defined by simple mapping algorithms. 

In this paper, we present simple mapping algorithms defining DIMs of all odd length. It turns out that we need separate mapping algorithms for DIMs of length $n\equiv1\mod 4$ and DIMs of length $n\equiv3\mod 4$. That these algorithms indeed define DIMs is proved by a straightforward method. This method also provides us a proof that the DIMs of even length in \cite{kwankyu:2005} indeed defines DIMs. Since the proof is easier and direct than the original proof, we include it in the next section.

We introduce some convenient notations. For a set $S$, $\#S$ denotes the number of elements in the set. For a predicate $P$, $[P]$ gives $1$ or $0$ if $P$ is true or false, respectively. For integers $u$ and $v$, $\gd(u,v)$ gives $1$ if $u=v$ and $0$ otherwise.

\section{DIMs of even length}

The following lemma is easy but very useful in subsequent proofs.
 \begin{lem}\label{lem_abc}
Let $a_1,a_2,\dots,a_n$ be a sequence with $a_i$ either $0$ or $1$ and $n\ge2$.  Then
\begin{align}
	\begin{split}\label{equ_dsqqq}
	&a_1a_2+a_2a_3+\dots+a_{n-1}a_n\\
	&\quad=\#\set{1\le i\le n\mid a_i=1}-B(a_i\mid 1\le i\le n)\\
	&\quad\le\#\set{1\le i\le n\mid a_i=1}-[\text{$a_i=1$ for some $1\le i\le n$}],
	\end{split}\\
	\begin{split}\label{equ_dxjza}
	&a_1a_2+a_2a_3+\dots+a_{n-1}a_n+a_na_1\\
	&\quad\le\#\set{1\le i\le n\mid a_i=1}\\
	&\qquad-[\text{$a_i=1$ and $a_j=0$ for some $1\le i,j\le n$}],	
	\end{split}
\end{align}
where $B(a_i\mid 1\le i\le n)$ denotes the number of blocks of consecutive $1$'s in the sequence. 
\end{lem}

\begin{proof}
First assertion is obvious if we notice that a block of consecutive $1$'s in the sequence contributes $l-1$ to the sum if the length of the block is $l$. To prove the second assertion, we apply the first to see
\[
\begin{split}
	&(a_1a_2+a_2a_3+\dots+a_{n-1}a_n)+a_na_1\\
	&\quad=\#\set{1\le i\le n\mid a_i=1}-B(a_i\mid 1\le i\le n)+a_na_1.
\end{split}
\]
Suppose $a_na_1=0$. Then the assertion follows since
\[
\begin{split}
	B(a_i\mid 1\le i\le n)&\ge[\text{$a_i=1$ for some $1\le i\le n$}]\\
	&=[\text{$a_i=1$ and $a_j=0$ for some $1\le i,j\le n$}].
\end{split}
\]
Suppose $a_na_1=1$. The claim also follows since
\[
\begin{split}
	B(a_i\mid 1\le i\le n)&>[\text{$a_j=0$ for some $1\le j\le n$}]\\
	&=[\text{$a_i=1$ and $a_j=0$ for some $1\le i,j\le n$}].	
\end{split}
\]
\end{proof}

\begin{figure}[h]
\begin{center}
\begin{minipage}[b]{\textwidth}
\begin{xy}
0;<4pt,0pt>:
,@={
0;a(0):(13,0)*=<1.6pc>[o][F]{2r}
,0;a(40):(13,0)*=<1.6pc>[o][F]{1}
,0;a(40):(13,0)*=<1.6pc>[o][F]{2}
,0;a(40):(13,0)*=<1.6pc>[o][F]{3}
,0;a(40):(13,0)*=<1.6pc>[o][F]{4}
,0;a(120):(13,0)*=<1.6pc>[o][F]{\scriptstyle 2r-2}
,0;a(40):(13,0)*=<1.6pc>[o][F]{\scriptstyle 2r-1}
}
,{\ar@{<->}s6;s5}?(.5)*+!L{u_{2r}}
,{\ar@{<->}s5;s4}?(.6)*+!LD{u_1}
,{\ar@{<->}s4;s3}?(.3)*+!RD{u_2}
,{\ar@{<->}s3;s2}?*+!RD{u_3}
,{\ar@{<->}s1;s0}?(.4)*+!LU{u_{2r-2}}
,{\ar@{<->}s0;s6}?(.4)*+!L{u_{2r-1}}
,0,{\ellipse(13.5)va(175)^,^,va(267)^{.}}
\end{xy}
\end{minipage}
\end{center}
\caption{\label{fig1} Swapping order of $z_{2r}$}
\end{figure}
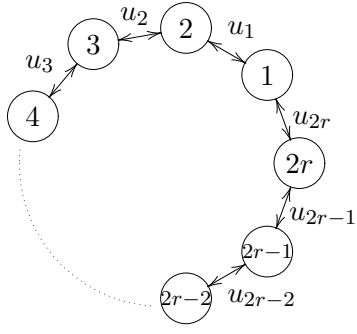

We now recall the DIMs of even length given in \cite{kwankyu:2005}. Let $r\ge2$. Let $z_{2r}$ be the map from $\Z_2^{2r}$ to $S_{2r}$ defined by
\begin{ttfamily}\bfseries
\begin{tabbing}ss\=ss\=ss\=\kill
    Mapping algorithm A \\
    Input: $u=(u_1,\dots,u_n)\in\Z_2^n$ $(n=2r)$ \\
    Output: $x=(x_1,\dots,x_n)\in S_n$ \\
    begin   \\
    \>  $(x_1,x_2,\dots,x_n)\leftarrow (1,2,\dots,n)$; \\
    \>  for $i$ from $1$ to $r$ do \\
    \>  \>  if $u_{2i-1}=1$ then $\mathrm{swap}(x_{2i-1},x_{2i})$;    \\
    \>  for $i$ from $1$ to $r$ do \\
    \>  \>  if $u_{2i}=1$ then $\mathrm{swap}(x_{2i},x_{2i+1})$; \\
    end
\end{tabbing}
\end{ttfamily}
It is shown in \cite{kwankyu:2005}, as a special case of a more general construction, that $z_{2r}$ are DIMs of length $2r$. We will give a direct proof of this shortly, using a method which we will also apply for DIMs of odd length. But first let us understand what this mapping algorithm does. Let $u$ be a binary vector. Consider the diagram in Figure~\ref{fig1}, which shows the permutation $(x_1,x_2,\dots,x_{2r})$ lying on a big circle with $x_i=i$ $(1\le i\le 2r)$ in small circles. Now each $u_i$, if it is $1$, swaps the two components in the small circles between which $u_i$ is placed, and this is done in the order
\[
	u_1\to u_3\to\dots\to u_{2r-1}\to u_2\to u_4\to\dots\to u_{2r}.
\]
The resulting permutation is what $u$ is mapped to under $z_{2r}$.

\begin{thm}
Let $r\ge2$. The maps $z_{2r}$ are DIMs of length $2r$.
\end{thm}

\begin{proof}
Let $n=2r$. Let $u$, $v$ be two distinct binary vectors, and let $z_{2r}(u)=x$ and $z_{2r}(v)=y$. We need to show $d(x,y)>d(u,v)$ if $d(u,v)<n$ and $d(x,y)=n$ if $d(u,v)=n$.
Let 
\[
	T=\set{1\le i\le n\mid\text{$x_j=y_j=i$ for some $1\le j\le n$}}.
\]
That is, $i\in T$ if and only if $i$'s in $x$ and $y$ are at the same position. Clearly $d(x,y)=n-\#T$ and $d(u,v)=n-\#\set{1\le i\le n\mid u_i=v_i}$. So we may rephrase our goal as 
\[
	\#T\le\#\set{1\le i\le n\mid\gd(u_i,v_i)=1}
\]
with equality only when the right side is zero, that is, when $d(u,v)=n$.

For each $1\le i\le n$, we can determine precise conditions to have $i\in T$. To give an example, let $i=1$, and view the diagram in Figure~\ref{fig1}. If $u_1\neq v_1$, then it is clear that the position of $1$ in $x$ will never be the same with that of $1$ in $y$ so that $1\notin T$. Suppose $u_1=v_1=0$, then $1$'s in $x$ and $y$ will be at the same position if and only if $u_{2r}=v_{2r}$. On the other hand, if $u_1=v_1=1$, then it is clear that $1\in T$ if and only if $u_2=v_2$. These cases exhaust all possibilities to have $1\in T$. Note that in this reasoning, the order in which swappings take place is crucial. We can express the result as follows: 
\begin{equation}\label{equ_cbaaj}
\begin{split}
    1\in T&\iff\gd(u_1,0)\gd(v_1,0)\gd(u_{2r},v_{2r})\\
    &\quad+\gd(u_1,1)\gd(v_1,1)\gd(u_2,v_2)=1.
\end{split}
\end{equation}
In a similar way, we obtain
\begin{align}
\begin{split}\label{equ_djccc}
    2\in T&\iff\gd(u_1,0)\gd(v_1,0)\gd(u_2,v_2)\\
    &\quad+\gd(u_1,1)\gd(v_1,1)\gd(u_{2r},v_{2r})=1,
\end{split}\\\notag
&\quad\vdots\\
\begin{split}\label{equ_wkwww}
    2r-1\in T&\iff\gd(u_{2r-1},0)\gd(v_{2r-1},0)\gd(u_{2r-2},v_{2r-2})\\
    &\quad+\gd(u_{2r-1},1)\gd(v_{2r-1},1)\gd(u_{2r},v_{2r})=1,
\end{split}\\
\begin{split}\label{equ_ffwdw}
    2r\in T&\iff\gd(u_{2r-1},0)\gd(v_{2r-1},0)\gd(u_{2r},v_{2r})\\
    &\quad+\gd(u_{2r-1},1)\gd(v_{2r-1},1)\gd(u_{2r-2},v_{2r-2})=1.
\end{split}
\end{align}
We may simplify by summing the expressions in right side in pairs like
\begin{align*}	\eqref{equ_cbaaj}+\eqref{equ_djccc}&=\gd(u_1,v_1)\gd(u_2,v_2)+\gd(u_1,v_1)\gd(u_{2r},v_{2r}),\\
&\vdots\\
\begin{split}
\eqref{equ_wkwww}+\eqref{equ_ffwdw}&=\gd(u_{2r-1},v_{2r-1})\gd(u_{2r-2},v_{2r-2})\\
&\quad+\gd(u_{2r-1},v_{2r-1})\gd(u_{2r},v_{2r}).
\end{split}
\end{align*}
So we obtain
\[
\begin{split}
\#T&=\eqref{equ_cbaaj}+\eqref{equ_djccc}+\dots+\eqref{equ_wkwww}+\eqref{equ_ffwdw}\\
	&=\gd(u_1,v_1)\gd(u_2,v_2)+\dots+\gd(u_{2r-1},v_{2r-1})\gd(u_{2r},v_{2r})\\
	&\quad+\gd(u_{2r},v_{2r})\gd(u_1,v_1)\\
	&\le\#\set{1\le i\le n\mid \gd(u_i,v_i)=1}\\
	&\quad-[\text{$\gd(u_i,v_i)=1$ and $\gd(u_j,v_j)=0$ for some $1\le i,j\le n$}]\\
	&=\#\set{1\le i\le n\mid \gd(u_i,v_i)=1}\\
	&\quad-[\text{$\gd(u_i,v_i)=1$ for some $1\le i,j\le n$}],
\end{split}
\]
where we used \eqref{equ_dxjza} of Lemma \ref{lem_abc}. Now our claim follows.
\end{proof}

\section{DIMs of odd length}

In this section, we present simple mapping algorithms defining DIMs of odd length. One algorithm defines DIMs of length $4r+1$ with $r\ge1$. The other defines DIMs of length $4r-1$ with $r\ge 2$. 

\subsection{DIMs of length $4r+1$}

Let $n=4r+1$ with $r\ge 1$. Let $z_{4r+1}$ be the map from $\Z_2^{4r+1}$ to $S_{4r+1}$ defined by 
\begin{ttfamily}\bfseries
\begin{tabbing}ss\=ss\=ss\=\kill
    Mapping algorithm B \\
    Input: $u=(u_1,\dots,u_n)\in\Z_2^n$ $(n=4r+1)$ \\
    Output: $x=(x_1,\dots,x_n)\in S_n$ \\
    begin   \\
    \>  $(x_1,x_2,\dots,x_n)\leftarrow (1,2,\dots,n)$; \\
    \>  for $i$ from $1$ to $2r$ do \\
    \>  \>  if $u_{2i-1}=1$ then $\mathrm{swap}(x_{2i-1},x_{2i})$;    \\
    \>  if $u_n=1$ then $\mathrm{swap}(x_n,x_1)$;    \\
    \>  if $u_n=1$ then $\mathrm{swap}(x_1,x_{2r+1})$;    \\
    \>  for $i$ from $1$ to $2r$ do \\
    \>  \>  if $u_{2i}=1$ then $\mathrm{swap}(x_{2i},x_{2i+1})$;    \\
    end
\end{tabbing}
\end{ttfamily}

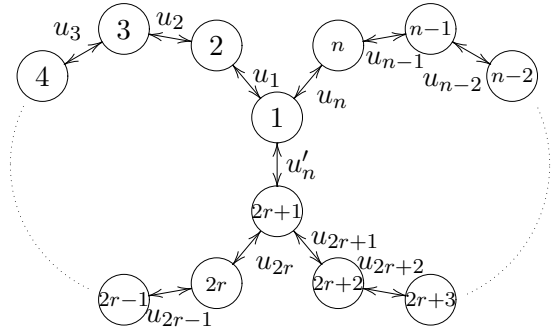
\begin{figure}[h]
\begin{center}
\begin{minipage}[b]{\textwidth}
\begin{xy}
0;<4pt,0pt>:
,@={
(0,0);a(-20):(13,0)*=<1.6pc>[o][F]{\scriptstyle 2r+1}
,(0,0);a(40):(13,0)*=<1.6pc>[o][F]{1}
,(0,0);a(40):(13,0)*=<1.6pc>[o][F]{2}
,(0,0);a(40):(13,0)*=<1.6pc>[o][F]{3}
,(0,0);a(40):(13,0)*=<1.6pc>[o][F]{4}
,(0,0);a(120):(13,0)*=<1.6pc>[o][F]{\scriptstyle 2r-1}
,(0,0);a(40):(13,0)*=<1.6pc>[o][F]{\scriptstyle 2r}
}
,{\ar@{<->}s6;s5}?(.5)*+!L{u_n'}
,{\ar@{<->}s5;s4}?(.55)*+!L{u_1}
,{\ar@{<->}s4;s3}?(.5)*+!D{u_2}
,{\ar@{<->}s3;s2}?(.4)*+!RD{u_3}
,{\ar@{<->}s1;s0}?(.6)*!U(3){u_{2r-1}}
,{\ar@{<->}s0;s6}?(.5)*+!LU{u_{2r}}
,(0,0),{\ellipse(13)va(155)^,^,va(-115)^{.}}
,(24.5,0);p+(-1,0):% Set new center
,@i@={
(0,0);a(-20):(13,0)*=<1.6pc>[o][F]i{1}%s6
,(0,0);a(40):(13,0)*=<1.6pc>[o][F]i{\scriptstyle 2r+1}%s5
,(0,0);a(40):(13,0)*=<1.6pc>[o][F]{\scriptstyle 2r+2}%s4
,(0,0);a(40):(13,0)*=<1.6pc>[o][F]{\scriptstyle 2r+3}%s3
,(0,0);a(120):(13,0)*=<1.6pc>[o][F]{\scriptstyle n-2}%s2
,(0,0);a(40):(13,0)*=<1.6pc>[o][F]{\scriptstyle n-1}%s1
,(0,0);a(40):(13,0)*=<1.6pc>[o][F]{\scriptstyle n}%s0
}
,{\ar@{<->}s5;s4}?(.4)*+!L{u_{2r+1}}
,{\ar@{<->}s4;s3}?(.57)*++!D{u_{2r+2}}
,{\ar@{<->}s2;s1}?(.25)*+!UR{u_{n-2}}
,{\ar@{<->}s1;s0}?(.8)*+!UL{u_{n-1}}
,{\ar@{<->}s0;s6}?(.5)*+!UL{u_{n}}
,(0,0),{\ellipse(13.5)va(115)^,^,va(205)^{.}}
\end{xy}
\end{minipage}
\end{center}
\caption{\label{fig2} Swapping order of $z_{4r+1}$}
\end{figure}

We give an intuitive description of what the algorithm does. Let $u$ be a binary vector given as input. The diagram in Figure~\ref{fig2} shows the permutation initialized with $x_i=i$ for $1\le i\le n$. Note that there are $2r+1$ components on the left big circle and $2r+2$ components on the right big circle and two components $x_1$ and $x_{2r+1}$ are shared between the two circles. Notice that there is a new variable $u_n'$ placed between $x_1=1$ and $x_{2r+1}=2r+1$. The variable $u_n'$ is simply set to the value of $u_n$. That is, $u_n'=u_n$.  Then swappings take place in the order
\begin{multline*}
u_1\to u_3\to\cdots\to u_{2r-1}\to u_{2r+1}\to\cdots\to u_{n-2}\to u_n \\
\to u_n'\to u_2\to\cdots\to u_{2r-2}\to u_{2r}\to u_{2r+2}\to\cdots\to u_{n-1}.
\end{multline*}
Now $u$ is mapped to the resulting permutation.
 
\begin{thm}
Let $r\ge 1$. The maps $z_{4r+1}$ are DIMs of length $4r+1$.
\end{thm}

\begin{proof}
Let $n=4r+1$. Let $z_{4r+1}(u)=x$ and $z_{4r+1}(v)=y$ for two distinct binary vectors $u$ and $v$. Let 
\[
	T=\set{1\le i\le n\mid\text{$x_j=y_j=i$ for some $1\le j\le n$}}.
\]
The proof proceeds in the same manner with that about DIMs of even length. So our goal is to show
\[
	\#T\le\#\set{1\le i\le n\mid\gd(u_i,v_i)=1}
\]
with equality only when the right side is zero.

It is a bit more complicated to determine the conditions to have $i\in T$. So we again give an example with $i=1$. In the following  reasoning, we must keep in mind the swapping order of $z_{4r+1}$. If $u_1\neq v_1$, then the position of $1$ in $x$ will never be the same with that in $y$, that is $1\notin T$. Suppose $u_1=v_1=1$, then $1$'s in $x$ and $y$ will be at the same position if and only if $u_2=v_2$. If $u_1=v_1=0$, then we need to consider $u_n$ and $v_n$. If $u_n\neq v_n$, then $1\notin T$. If $u_n=v_n=1$, then $1\in T$ if and only if $u_{n-1}=v_{n-1}$. If $u_n=v_n=0$, then we again consider $u_n'$ and $v_n'$. If $u_n'\neq v_n'$, then $1\notin T$. If $u_n'=v_n'=0$, then $1\in T$. If $u_n'=v_n'=1$, then we still consider $u_{2r}$ and $v_{2r}$ to see that $1\in T$ if and only if $u_{2r}=v_{2r}$.  We express the result formally as
\begin{equation}
\begin{split}\label{equ_jwjai}
    &1\in T\\
    &\iff\gd(u_1,0)\gd(v_1,0)\gd(u_n,0)\gd(v_n,0)\gd(u_n',0)\gd(v_n',0)\\
    &\quad+\gd(u_1,0)\gd(v_1,0)\gd(u_n,0)\gd(v_n,0)\gd(u_n',1)\gd(v_n',1)\gd(u_{2r},v_{2r})\\
    &\quad+\gd(u_1,0)\gd(v_1,0)\gd(u_n,1)\gd(v_n,1)\gd(u_{n-1},v_{n-1})\\
    &\quad+\gd(u_1,1)\gd(v_1,1)\gd(u_2,v_2)=1.
\end{split}
\end{equation}

Similar reasonings give
\begin{align}
\begin{split}\label{equ_ckwkz}
    &2\in T\\
    &\iff\gd(u_1,0)\gd(v_1,0)\gd(u_2,v_2)\\
    &\quad+\gd(u_1,1)\gd(v_1,1)\gd(u_n,0)\gd(v_n,0)\gd(u_n',0)\gd(v_n',0)\\
	&\quad+\gd(u_1,1)\gd(v_1,1)\gd(u_n,0)\gd(v_n,0)\gd(u_n',1)\gd(v_n',1)\gd(u_{2r},v_{2r})\\
    &\quad+\gd(u_1,1)\gd(v_1,1)\gd(u_n,1)\gd(v_n,1)\gd(u_{n-1},v_{n-1})=1,
\end{split}\\
\begin{split}\label{equ_xkccs}
    &3\in T\\
    &\iff\gd(u_3,0)\gd(v_3,0)\gd(u_2,v_2)\\
        &\quad+\gd(u_3,1)\gd(v_3,1)\gd(u_4,v_4)=1,
\end{split}\\
\begin{split}\label{equ_kaskl}
    &4\in T\\
    &\iff\gd(u_3,0)\gd(v_3,0)\gd(u_4,v_4)\\
        &\quad+\gd(u_3,1)\gd(v_3,1)\gd(u_2,v_2)=1,
\end{split}\\\notag
&\quad\vdots\\
\begin{split}\label{equ_qwdjd}
    &2r-1\in T\\
    &\iff\gd(u_{2r-1},0)\gd(v_{2r-1},0)\gd(u_{2r-2},v_{2r-2})\\
        &\quad+\gd(u_{2r-1},1)\gd(v_{2r-1},1)\gd(u_{2r},v_{2r})=1,
\end{split}\\
\begin{split}\label{equ_cmsmw}
    &2r\in T\\
    &\iff\gd(u_{2r-1},0)\gd(v_{2r-1},0)\gd(u_{2r},v_{2r})\\
        &\quad+\gd(u_{2r-1},1)\gd(v_{2r-1},1)\gd(u_{2r-2},v_{2r-2})=1,
\end{split}\\
\begin{split}\label{equ_dkwua}
   	&2r+1\in T\\
   	&\iff\gd(u_{2r+1},0)\gd(v_{2r+1},0)\gd(u_n',0)\gd(v_n',0)\gd(u_{2r},v_{2r})\\
   	&\quad+\gd(u_{2r+1},0)\gd(v_{2r+1},0)\gd(u_n',1)\gd(v_n',1)\\
   	&\quad+\gd(u_{2r+1},1)\gd(v_{2r+1},1)\gd(u_{2r+2},v_{2r+2})=1,
\end{split}\\
\begin{split}\label{equ_fwkzs}
   	&2r+2\in T\\
   	&\iff\gd(u_{2r+1},0)\gd(v_{2r+1},0)\gd(u_{2r+2},v_{2r+2})\\
   	&\quad+\gd(u_{2r+1},1)\gd(v_{2r+1},1)\gd(u_n',0)\gd(v_n',0)\gd(u_{2r},v_{2r})\\
   	&\quad+\gd(u_{2r+1},1)\gd(v_{2r+1},1)\gd(u_n',1)\gd(v_n',1)=1,
\end{split}\\
\begin{split}\label{equ_qowkx}
    &2r+3\in T\\
    &\iff\gd(u_{2r+3},0)\gd(v_{2r+3},0)\gd(u_{2r+2},v_{2r+2})\\
        &\quad+\gd(u_{2r+3},1)\gd(v_{2r+3},1)\gd(u_{2r+4},v_{2r+4})=1,
\end{split}\\\notag
&\quad\vdots\\
\begin{split}\label{equ_dkjac}
    &n-1\in T\\
    &\iff\gd(u_{n-2},0)\gd(v_{n-2},0)\gd(u_{n-1},v_{n-1})\\
        &\quad+\gd(u_{n-2},1)\gd(v_{n-2},1)\gd(u_{n-3},v_{n-3})=1,
\end{split}\\
\begin{split}\label{equ_cjwja}
    &n\in T\\
    &\iff\gd(u_n,0)\gd(v_n,0)\gd(u_{n-1},v_{n-1})\\
       &\quad+\gd(u_n,1)\gd(v_n,1)\gd(u_n',0)\gd(v_n',0)\\
       &\quad+\gd(u_n,1)\gd(v_n,1)\gd(u_n',1)\gd(v_n',1)\gd(u_{2r},v_{2r})=1.
\end{split}
\end{align}
Note that \eqref{equ_xkccs}, \eqref{equ_kaskl}, \dots, \eqref{equ_qwdjd}, \eqref{equ_cmsmw} and \eqref{equ_qowkx}, \dots, \eqref{equ_dkjac} do not appear if $r=1$. 

Now $\#T$ equals the total of sums in the right sides of \eqref{equ_jwjai}--\eqref{equ_cjwja}. The sum is simplified greatly if we first compute in groups. Thus
\begin{align*}
	\eqref{equ_jwjai}+\eqref{equ_ckwkz}
	&=\gd(u_1,v_1)\gd(v_2,v_2)+\gd(u_1,v_1)\gd(u_n,0)\gd(v_n,0)\\
	&\quad+\gd(u_1,v_1)\gd(u_{n-1},v_{n-1})\gd(u_n,1)\gd(v_n,1),\\
	\eqref{equ_xkccs}+\dots+\eqref{equ_cmsmw}	
	&=\gd(u_2,v_2)\gd(u_3,v_3)
	+\gd(u_3,v_3)\gd(u_4,v_4)+\cdots\\
	&\quad+\gd(u_{2r-1},v_{2r-1})\gd(u_{2r},v_{2r}),\\				
	\eqref{equ_dkwua}+\eqref{equ_fwkzs}
	&=\gd(u_{2r+1},v_{2r+1})\gd(u_{2r+2},v_{2r+2})\\
	&\quad+\gd(u_{2r+1},v_{2r+1})\gd(u_{2r},v_{2r})\gd(u_n,0)\gd(v_n,0)\\
	&\quad+\gd(u_{2r+1},v_{2r+1})\gd(u_n,1)\gd(v_n,1),\\				
	\eqref{equ_qowkx}+\dots+\eqref{equ_dkjac}
	&=\gd(u_{2r+2},v_{2r+2})\gd(u_{2r+3},v_{2r+3})
	+\cdots\\
	&\quad+\gd(u_{n-2},v_{n-2})\gd(u_{n-1},v_{n-1}),\\	
	\eqref{equ_cjwja}
	&=\gd(u_{n-1},v_{n-1})\gd(u_n,0)\gd(v_n,0)\\
	&\quad+\gd(u_{2r},v_{2r})\gd(u_n,1)\gd(v_n,1).
\end{align*}

Summing all, $\#T$ equals
\[
\begin{split}
	&\gd(u_1,v_1)\gd(v_2,v_2)+\dots+\gd(u_{2r-1},v_{2r-1})\gd(u_{2r},v_{2r})\\
	&+\gd(u_{2r+1},v_{2r+1})\gd(u_{2r+2},v_{2r+2})
	+\dots+\gd(u_{n-2},v_{n-2})\gd(u_{n-1},v_{n-1})\\
	&+\gd(u_1,v_1)\gd(u_n,0)\gd(v_n,0)
	+\gd(u_1,v_1)\gd(u_{n-1},v_{n-1})\gd(u_n,1)\gd(v_n,1)\\
	&+\gd(u_{2r},v_{2r})\gd(u_{2r+1},v_{2r+1})\gd(u_n,0)\gd(v_n,0)\\
	&+\gd(u_{2r+1},v_{2r+1})\gd(u_n,1)\gd(v_n,1)\\
	&+\gd(u_{n-1},v_{n-1})\gd(u_n,0)\gd(v_n,0)+\gd(u_{2r},v_{2r})\gd(u_n,1)\gd(v_n,1).
\end{split}
\]

We now treat three different cases. For the case $u_n=0$ and $v_n=0$:
\[
\begin{split}
	\#T&=\gd(u_1,v_1)\gd(u_2,v_2)+\dots+\gd(u_{n-2},v_{n-2})\gd(u_{n-1},v_{n-1})\\
	&\quad+\gd(u_1,v_1)+\gd(u_{n-1},v_{n-1})\\
	&\le\#\set{1\le i\le n-1\mid \gd(u_i,v_i)=1}-B(\gd(u_i,v_i)\mid 1\le i\le n-1)\\
	&\quad+\gd(u_1,v_1)+\gd(u_{n-1},v_{n-1})\\
	&=\#\set{1\le i\le n\mid \gd(u_i,v_i)=1}-1-B(\gd(u_i,v_i)\mid 1\le i\le n-1)\\
	&\quad+\gd(u_1,v_1)+\gd(u_{n-1},v_{n-1})\\
	&\le\#\set{1\le i\le n\mid \gd(u_i,v_i)=1}-1,
\end{split}
\]
where \eqref{equ_dsqqq} of Lemma \ref{lem_abc} is used. The last inequality is valid since 
\[
	B(\gd(u_i,v_i)\mid 1\le i\le n-1)\ge\gd(u_1,v_1)+\gd(u_{n-1},v_{n-1}),
\]
which is itself easy to verify. If at most one of $\gd(u_1,v_1)=1$ and $\gd(u_{n-1},v_{n-1})=1$ is true, then the inequality clearly holds. If $\gd(u_1,v_1)=\gd(u_{n-1},v_{n-1})=1$, then there must be some $1<j<n-1$ such that $\gd(u_j,v_j)=0$ because $u$ and $v$ are distinct binary vectors, and hence $B(\gd(u_i,v_i)\mid 1\le i\le n-1)\ge 2$.

For the case $u_n=1$ and $v_n=1$:
\[
\begin{split}
	\#T&=\gd(u_1,v_1)\gd(u_2,v_2)+\dots+\gd(u_{2r-1},v_{2r-1})\gd(u_{2r},v_{2r})\\
	&\quad+\gd(u_{2r+1},v_{2r+1})\gd(u_{2r+2},v_{2r+2})
	+\cdots\\
	&\quad+\gd(u_{n-2},v_{n-2})\gd(u_{n-1},v_{n-1})\\
	&\quad+\gd(u_1,v_1)\gd(u_{n-1},v_{n-1})+\gd(u_{2r},v_{2r})+\gd(u_{2r+1},v_{2r+1})\\
	&\le\#\set{1\le i\le 2r\mid \gd(u_i,v_i)=1}-B(\gd(u_i,v_i)\mid 1\le i\le 2r)\\
	&\quad+\#\set{2r+1\le i\le n-1\mid\gd(u_i,v_i)=1}\\
	&\quad-B(\gd(u_i,v_i)\mid2r+1\le i\le n-1)\\
	&\quad+\gd(u_1,v_1)\gd(u_{n-1},v_{n-1})+\gd(u_{2r},v_{2r})+\gd(u_{2r+1},v_{2r+1})\\
	&=\#\set{1\le i\le n\mid \gd(u_i,v_i)=1}-1-B(\gd(u_i,v_i)\mid 1\le i\le 2r)\\
	&\quad-B(\gd(u_i,v_i)\mid2r+1\le i\le n-1)\\
	&\quad+\gd(u_1,v_1)\gd(u_{n-1},v_{n-1})+\gd(u_{2r},v_{2r})+\gd(u_{2r+1},v_{2r+1})\\			&\le\#\set{1\le i\le n\mid \gd(u_i,v_i)=1}-1.
\end{split}
\]
The last inequality is valid because 
\[
\begin{split}
	B(\gd(u_i,v_i)\mid 1\le i\le 2r)+B(\gd(u_i,v_i)\mid2r+1\le i\le n-1)\\
	\ge\gd(u_1,v_1)\gd(u_{n-1},v_{n-1})+\gd(u_{2r},v_{2r})+\gd(u_{2r+1},v_{2r+1}),
\end{split}
\]
which is easily verified in a similar way as above.

For the case $u_n\neq v_n$:
\[
\begin{split}
	\#T&=\gd(u_1,v_1)\gd(u_2,v_2)+\dots+\gd(u_{2r-1},v_{2r-1})\gd(u_{2r},v_{2r})\\
	&\quad+\gd(u_{2r+1},v_{2r+1})\gd(u_{2r+2},v_{2r+2})
	+\cdots\\
	&\quad+\gd(u_{n-2},v_{n-2})\gd(u_{n-1},v_{n-1})\\
	&\le\#\set{1\le i\le 2r\mid \gd(u_i,v_i)=1}-B(\gd(u_i,v_i)\mid 1\le i\le 2r)\\
	&\quad+\#\set{2r+1\le i\le n-1\mid\gd(u_i,v_i)=1}\\
	&\quad-B(\gd(u_i,v_i)\mid2r+1\le i\le n-1)\\
	&=\#\set{1\le i\le n\mid \gd(u_i,v_i)=1}-B(\gd(u_i,v_i)\mid 1\le i\le 2r)\\
	&\quad-B(\gd(u_i,v_i)\mid2r+1\le i\le n-1)\\
	&\le\#\set{1\le i\le n\mid \gd(u_i,v_i)=1}\\
	&\quad-[\text{$\gd(u_i,v_i)=1$ for some $1\le i\le n$}],
\end{split}
\]
where the last inequality is easily verified.

Now the three cases combine to give the required result.
\end{proof}

\subsection{DIMs of length $4r-1$}

We now turn to DIMs of length $4r-1$. Much of the material in this subsection goes parallel with that of the previous subsection. So we omit some repetitive details. Let $n=4r-1$ with $r\ge 2$. Let $z_{4r-1}$ be the map from $\Z_2^{4r-1}$ to $S_{4r-1}$ defined by 
\begin{ttfamily}\bfseries
\begin{tabbing}ss\=ss\=ss\=\kill
    Mapping algorithm C \\
    Input: $u=(u_1,\dots,u_n)\in\Z_2^n$ $(n=4r-1)$ \\
    Output: $x=(x_1,\dots,x_n)\in S_n$ \\
    begin   \\
    \>  $(x_1,x_2,\dots,x_n)\leftarrow (1,2,\dots,n)$; \\
    \>  for $i$ from $1$ to $2r-1$ do \\
    \>  \>  if $u_{2i-1}=1$ then $\mathrm{swap}(x_{2i-1},x_{2i})$;    \\
    \>  if $u_n=1$ then $\mathrm{swap}(x_n,x_{2r})$;    \\
    \>  if $u_{2r}=1$ then $\mathrm{swap}(x_1,x_{2r})$;    \\
    \>  for $i$ from $1$ to $2r-1$ do \\
    \>  \>  if $u_{2i}=1$ then $\mathrm{swap}(x_{2i},x_{2i+1})$;    \\
    end
\end{tabbing}
\end{ttfamily}

\begin{figure}[h]
\begin{center}
\begin{minipage}[b]{\textwidth}
\begin{xy}
0;<4pt,0pt>:
,@={
0;a(0):(16,0)*=<1.6pc>[o][F]{2r}
,0;a(40):(13,0)*=<1.6pc>[o][F]{1}
,0;a(40):(13,0)*=<1.6pc>[o][F]{2}
,0;a(40):(13,0)*=<1.6pc>[o][F]{3}
,0;a(40):(13,0)*=<1.6pc>[o][F]{4}
,0;a(120):(13,0)*=<1.6pc>[o][F]{\scriptstyle 2r-2}
,0;a(40):(13,0)*=<1.6pc>[o][F]{\scriptstyle 2r-1}
}
,{\ar@{<->}s6;s5}?(.5)*!LD{u_{2r}'}
,{\ar@{<->}s5;s4}?(.6)*+!LD{u_1}
,{\ar@{<->}s4;s3}?(.4)*+!RD{u_2}
,{\ar@{<->}s3;s2}?*+!RD{u_3}
,{\ar@{<->}s1;s0}?(.4)*+!LU{u_{2r-2}}
,{\ar@{<->}s0;s6}?(.5)*!LU{u_{2r-1}}
,0,{\ellipse(13.5)va(175)^,^,va(267)^{.}}
,(32,0);p+(-1,0):(0,0)% Set new center
,@i@={
(0,0);a(0):(16,0)*=<1.6pc>[o][F]i{2r}%s6
,(0,0);a(40):(13,0)*=<1.6pc>[o][F]{n}%s5
,(0,0);a(40):(13,0)*=<1.6pc>[o][F]{\scriptstyle n-1}%s4
,(0,0);a(40):(13,0)*=<1.6pc>[o][F]{\scriptstyle n-2}%s3
,(0,0);a(120):(13,0)*=<1.6pc>[o][F]{\scriptstyle 2r+3}%s2
,(0,0);a(40):(13,0)*=<1.6pc>[o][F]{\scriptstyle 2r+2}%s1
,(0,0);a(40):(13,0)*=<1.6pc>[o][F]{\scriptstyle 2r+1}%s0
}
,{\ar@{<->}s6;s5}?(.45)*+!L{u_n}
,{\ar@{<->}s5;s4}?(.3)*+!DL{u_{n-1}}
,{\ar@{<->}s4;s3}?(.4)*++!D{u_{n-2}}
,{\ar@{<->}s2;s1}?(.55)*++!U{u_{2r+2}}
,{\ar@{<->}s1;s0}?(.7)*+!UL{u_{2r+1}}
,{\ar@{<->}s0;s6}?(.5)*+!UL{u_{2r}}
,(0,0),{\ellipse(13.5)va(135)^,^,va(227)^{.}}
\end{xy}
\end{minipage}
\end{center}
\caption{\label{fig3} Swapping order of $z_{4r-1}$}
\end{figure}
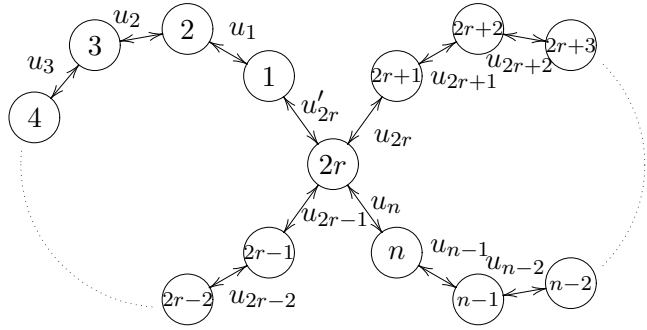

We again give an intuitive description of what the algorithm does. Let $u$ be a binary vector given as input. The diagram in Figure~\ref{fig3} shows the permutation initialized with $x_i=i$ for $1\le i\le 4r-1$. Note that there are $2r$ elements on the left big circle and $2r$ elements on the right big circle and that $x_{2r}=2r$ is shared between the two. Notice that there is a new variable $u_{2r}'$ between $x_1=1$ and $x_{2r}=2r$. The variable $u_{2r}'$ is set to the value of $u_{2r}$ so that $u_{2r}'=u_{2r}$. Then swappings take place in the order
\begin{multline*}
u_1\to u_3\to\cdots\to u_{2r-1}\to u_{2r+1}\to\cdots\to u_{n-2}\to u_n \\
\to u_{2r}'\to u_2\to\cdots\to u_{2r-2}\to u_{2r}\to u_{2r+2}\to\cdots\to u_{n-1}.
\end{multline*}
Then $u$ is mapped to the resulting permutation under $z_{4r-1}$.

\begin{thm}
Let $r\ge2$. The maps $z_{4r-1}$ are DIMs of length $4r-1$.
\end{thm}

\begin{proof}
Let $n=4r-1$. Let $z_{4r-1}(u)=x$ and $z_{4r-1}(v)=y$ for two distinct binary vectors $u$ and $v$. 
Let 
\[
	T=\set{1\le i\le n\mid\text{$x_j=y_j=i$ for some $1\le j\le n$}}.
\] 
Again our goal is to show
\[
	\#T\le\#\set{1\le i\le n\mid\gd(u_i,v_i)=1}
\]
with equality only when the right side is zero.

The conditions to have $i\in T$ for $1\le i\le n$ are as follows.
\begin{align}
\begin{split}\label{equ_abchw}
    &1\in T\\
    &\iff\gd(u_1,0)\gd(v_1,0)\gd(u_{2r}',0)\gd(v_{2r}',0)\\
        &\quad+\gd(u_1,0)\gd(v_1,0)\gd(u_{2r}',1)\gd(v_{2r}',1)\gd(u_{2r},v_{2r})\\
        &\quad+\gd(u_1,1)\gd(v_1,1)\gd(u_2,v_2)=1,
\end{split}\\
\begin{split}\label{equ_ckwkc}
   &2\in T\\
    &\iff\gd(u_1,0)\gd(v_1,0)\gd(u_2,v_2)\\
        &\quad+\gd(u_1,1)\gd(v_1,1)\gd(u_{2r}',0)\gd(v_{2r}',0)\\
        &\quad+\gd(u_1,1)\gd(v_1,1)\gd(u_{2r}',1)\gd(v_{2r}',1)\gd(u_{2r},v_{2r})=1,
\end{split}\\
\begin{split}\label{equ_sjcjq}
    &3\in T\\
    &\iff\gd(u_3,0)\gd(v_3,0)\gd(u_2,v_2)\\
        &\quad+\gd(u_3,1)\gd(v_3,1)\gd(u_4,v_4)=1,
\end{split}\\
\begin{split}\label{equ_cjdwd}
    &4\in T\\
    &\iff\gd(u_3,0)\gd(v_3,0)\gd(u_4,v_4)\\
        &\quad+\gd(u_3,1)\gd(v_3,1)\gd(u_2,v_2)=1,
\end{split}\\\notag
&\quad\vdots\\
\begin{split}\label{equ_qkcix}
    &2r-2\in T\\
    &\iff\gd(u_{2r-3},0)\gd(v_{2r-3},0)\gd(u_{2r-2},v_{2r-2})\\
        &\quad+\gd(u_{2r-3},1)\gd(v_{2r-3},1)\gd(u_{2r-4},v_{2r-4})=1,
\end{split}\\
\begin{split}\label{equ_cjmwp}
   	&2r-1\in T\\
   	&\iff\gd(u_{2r-1},0)\gd(v_{2r-1},0)\gd(u_{2r-2},v_{2r-2})\\
    &\quad+\gd(u_{2r-1},1)\gd(v_{2r-1},1)\gd(u_n,0)\gd(v_n,0)
    \gd(u_{2r}',0)\gd(v_{2r}',0)\gd(u_{2r},v_{2r})\\
    &\quad+\gd(u_{2r-1},1)\gd(v_{2r-1},1)\gd(u_n,0)\gd(v_n,0)
    \gd(u_{2r}',1)\gd(v_{2r}',1)\\
    &\quad+\gd(u_{2r-1},1)\gd(v_{2r-1},1)\gd(u_n,1)\gd(v_n,1)
    \gd(u_{n-1},v_{n-1})=1,
\end{split}\\
\begin{split}\label{equ_cjkqd}
    &2r\in T\\
    &\iff\gd(u_{2r-1},1)\gd(v_{2r-1},1)\gd(u_{2r-2},v_{2r-2})\\
    &\quad+\gd(u_{2r-1},0)\gd(v_{2r-1},0)\gd(u_n,0)\gd(v_n,0)
    \gd(u_{2r}',0)\gd(v_{2r}',0)\gd(u_{2r},v_{2r})\\
    &\quad+\gd(u_{2r-1},0)\gd(v_{2r-1},0)\gd(u_n,0)\gd(v_n,0)
    \gd(u_{2r}',1)\gd(v_{2r}',1)\\
    &\quad+\gd(u_{2r-1},0)\gd(v_{2r-1},0)\gd(u_n,1)\gd(v_n,1)
    \gd(u_{n-1},v_{n-1})=1,
\end{split}\\
\begin{split}\label{equ_fkwpx}
    &2r+1\in T\\
    &\iff\gd(u_{2r+1},0)\gd(v_{2r+1},0)\gd(u_{2r},v_{2r})\\
        &\quad+\gd(u_{2r+1},1)\gd(v_{2r+1},1)\gd(u_{2r+2},v_{2r+2})=1,
\end{split}\\
\begin{split}\label{equ_axkdx}
    &2r+2\in T\\
    &\iff\gd(u_{2r+1},0)\gd(v_{2r+1},0)\gd(u_{2r+2},v_{2r+2})\\
        &\quad+\gd(u_{2r+1},1)\gd(v_{2r+1},1)\gd(u_{2r},v_{2r})=1,
\end{split}\\
\begin{split}\label{equ_xjqoi}
    &2r+3\in T\\
    &\iff\gd(u_{2r+3},0)\gd(v_{2r+3},0)\gd(u_{2r+2},v_{2r+2})\\
        &\quad+\gd(u_{2r+3},1)\gd(v_{2r+3},1)\gd(u_{2r+4},v_{2r+4})=1,
\end{split}\\\notag
&\quad\vdots\\
\begin{split}\label{equ_dkoad}
    &n-2\in T\\
    &\iff\gd(u_{n-2},0)\gd(v_{n-2},0)\gd(u_{n-3},v_{n-3})\\
        &\quad+\gd(u_{n-2},1)\gd(v_{n-2},1)\gd(u_{n-1},v_{n-1})=1,
\end{split}\\
\begin{split}\label{equ_cjajw}
    &n-1\in T\\
    &\iff\gd(u_{n-2},0)\gd(v_{n-2},0)\gd(u_{n-1},v_{n-1})\\
        &\quad+\gd(u_{n-2},1)\gd(v_{n-2},1)\gd(u_{n-3},v_{n-3})=1,
\end{split}\\
\begin{split}\label{equ_apskw}
    &n\in T\\
    &\iff\gd(u_n,0)\gd(v_n,0)\gd(u_{n-1},v_{n-1})\\
       &\quad+\gd(u_n,1)\gd(v_n,1)\gd(u_{2r}',0)\gd(v_{2r}',0)\gd(u_{2r},v_{2r})\\
       &\quad+\gd(u_n,1)\gd(v_n,1)\gd(u_{2r}',1)\gd(v_{2r}',1)=1.
\end{split}
\end{align}

Note that \eqref{equ_sjcjq}, \eqref{equ_cjdwd}, \dots, \eqref{equ_qkcix} and \eqref{equ_axkdx}, \eqref{equ_xjqoi}, \dots, \eqref{equ_dkoad} do not appear if $r=2$. We simplify by computing the sums in groups:
\begin{align*}
	\eqref{equ_abchw}+\eqref{equ_ckwkc}
	&=\gd(u_1,v_1)\gd(v_{2r},v_{2r})+\gd(u_1,v_1)\gd(v_2,v_2),\\
	\eqref{equ_sjcjq}+\dots+\eqref{equ_qkcix}	
	&=\gd(u_2,v_2)\gd(u_3,v_3)+\gd(u_3,v_3)\gd(u_4,v_4)+\cdots\\
	&\quad+\gd(u_{2r-3},v_{2r-3})\gd(u_{2r-2},v_{2r-2}),\\				
	\eqref{equ_cjmwp}+\eqref{equ_cjkqd}
	&=\gd(u_{2r-1},v_{2r-1})\gd(u_{2r-2},v_{2r-2})\\
	&\quad+\gd(u_{2r-1},v_{2r-1})\gd(u_{2r},v_{2r})\gd(u_n,0)\gd(v_n,0)\\
	&\quad+\gd(u_{2r-1},v_{2r-1})\gd(u_{n-1},v_{n-1})\gd(u_n,0)\gd(v_n,0),\\	
	\eqref{equ_fkwpx}+\dots+\eqref{equ_dkoad}
	&=\gd(u_{2r},v_{2r})\gd(u_{2r+1},v_{2r+1})\\
	&\quad+\gd(u_{2r+1},v_{2r+1})\gd(u_{2r+2},v_{2r+2})+\cdots\\
	&\quad+\gd(u_{n-2},v_{n-2})\gd(u_{n-1},v_{n-1}),\\	
	\eqref{equ_cjajw}+\eqref{equ_apskw}
	&=\gd(u_{n-1},v_{n-1})\gd(u_n,0)\gd(v_n,0)\\
	&\quad+\gd(u_{2r},v_{2r})\gd(u_n,1)\gd(v_n,1).
\end{align*}

Thus we obtain
\[
\begin{split}
	\#T&=\gd(u_1,v_1)\gd(v_{2r},v_{2r})\\
	&\quad+\gd(u_1,v_1)\gd(v_2,v_2)+\gd(u_2,v_2)\gd(u_3,v_3)\\
	&\quad+\gd(u_3,v_3)\gd(u_4,v_4)+\dots+\gd(u_{2r-2},v_{2r-2})\gd(u_{2r-1},v_{2r-1})\\
	&\quad+\gd(u_{2r-1},v_{2r-1})\gd(u_{2r},v_{2r})\gd(u_n,0)\gd(v_n,0)\\
	&\quad+\gd(u_{2r-1},v_{2r-1})\gd(u_{n-1},v_{n-1})\gd(u_n,0)\gd(v_n,0)\\
	&\quad+\gd(u_{2r},v_{2r})\gd(u_{2r+1},v_{2r+1})\\
	&\quad+\gd(u_{2r+1},v_{2r+1})\gd(u_{2r+2},v_{2r+2})+\cdots\\
	&\quad+\gd(u_{n-2},v_{n-2})\gd(u_{n-1},v_{n-1})\\
	&\quad+\gd(u_{n-1},v_{n-1})\gd(u_n,0)\gd(v_n,0)\\
	&\quad+\gd(u_{2r},v_{2r})\gd(u_n,1)\gd(v_n,1).
\end{split}
\]
We treat three separate cases. For the case $u_n=0$ and $v_n=0$:
\[
\begin{split}
	\#T&=\gd(u_1,v_1)\gd(u_2,v_2)+\dots+\gd(u_{2r},v_{2r})\gd(u_1,v_1)\\
	&\quad+\gd(u_{2r},v_{2r})\gd(u_{2r+1},v_{2r+1})+\dots\\
	&\quad+\gd(u_{n-2},v_{n-2})\gd(u_{n-1},v_{n-1})\\
	&\quad+\gd(u_{2r-1},v_{2r-1})\gd(u_{n-1},v_{n-1})+\gd(u_{n-1},v_{n-1})\\
	&\le\#\set{1\le i\le 2r\mid \gd(u_i,v_i)=1}\\
	&\quad-[\text{$\gd(u_i,v_i)=1$ and $\gd(u_j,v_j)=0$ for some $1\le i,j\le 2r$}]\\
	&\quad+\#\set{2r\le i\le n-1\mid\gd(u_i,v_i)=1}\\
	&\quad-[\text{$\gd(u_i,v_i)=1$ for some $2r\le i\le n-1$}]\\
	&\quad+\gd(u_{2r-1},v_{2r-1})\gd(u_{n-1},v_{n-1})+\gd(u_{n-1},v_{n-1})\\
	&=\#\set{1\le i\le n\mid \gd(u_i,v_i)=1}-\gd(u_{2r},v_{2r})-1\\
	&\quad-[\text{$\gd(u_i,v_i)=1$ and $\gd(u_j,v_j)=0$ for some $1\le i,j\le 2r$}]\\
	&\quad-[\text{$\gd(u_i,v_i)=1$ for some $2r\le i\le n-1$}]\\
	&\quad+\gd(u_{2r-1},v_{2r-1})\gd(u_{n-1},v_{n-1})+\gd(u_{n-1},v_{n-1})\\
	&\le\#\set{1\le i\le n\mid \gd(u_i,v_i)=1}-1\,\
\end{split}
\]
where the last inequality follows from
\[
\begin{split}
	&[\text{$\gd(u_i,v_i)=1$ and $\gd(u_j,v_j)=0$ for some $1\le i,j\le 2r$}]\\
	&\quad+[\text{$\gd(u_i,v_i)=1$ for some $2r\le i\le n-1$}]+\gd(u_{2r},v_{2r})\\
	&\ge\gd(u_{2r-1},v_{2r-1})\gd(u_{n-1},v_{n-1})
	+\gd(u_{n-1},v_{n-1}),
\end{split}
\]
which can be verified as follows. We only consider the case that $\gd(u_{2r-1},v_{2r-1})=\gd(u_{n-1},v_{n-1})=1$ as the other cases are trivial. Suppose this case. Then it suffices to note that $\gd(u_{2r},v_{2r})=0$ implies that 
\[
	[\text{$\gd(u_i,v_i)=1$ and $\gd(u_j,v_j)=0$ for some $1\le i,j\le 2r$}]=1.
\]

For the case $u_n=1$ and $v_n=1$:
\[
\begin{split}
	\#T&=\gd(u_1,v_1)\gd(u_2,v_2)+\cdots\\
	&\quad+\gd(u_{2r-2},v_{2r-2})\gd(u_{2r-1},v_{2r-1})\\
	&\quad+\gd(u_{2r},v_{2r})\gd(u_{2r+1},v_{2r+1})+\cdots\\
	&\quad+\gd(u_{n-2},v_{n-2})\gd(u_{n-1},v_{n-1})\\
	&\quad+\gd(u_1,v_1)\gd(u_{2r},v_{2r})+\gd(u_{2r},v_{2r})\\
	&\le\#\set{1\le i\le 2r-1\mid \gd(u_i,v_i)=1}\\
	&\quad-[\text{$\gd(u_i,v_i)=1$ for some $1\le i\le 2r-1$}]\\
	&\quad+\#\set{2r\le i\le n-1\mid\gd(u_i,v_i)=1}\\
	&\quad-[\text{$\gd(u_i,v_i)=1$ for some $2r\le i\le n-1$}]\\
	&\quad+\gd(u_1,v_1)\gd(u_{2r},v_{2r})+\gd(u_{2r},v_{2r})\\
	&=\#\set{1\le i\le n\mid \gd(u_i,v_i)=1}-1\\
	&\quad-[\text{$\gd(u_i,v_i)=1$ for some $1\le i\le 2r-1$}]\\
	&\quad-[\text{$\gd(u_i,v_i)=1$ for some $2r\le i\le n-1$}]\\
	&\quad+\gd(u_1,v_1)\gd(u_{2r},v_{2r})+\gd(u_{2r},v_{2r})\\
	&\le\#\set{1\le i\le n\mid \gd(u_i,v_i)=1}-1,
\end{split}
\]
where the last inequality follows from
\[
\begin{split}
	&[\text{$\gd(u_i,v_i)=1$ for some $1\le i\le 2r-1$}]\\
	&\quad+[\text{$\gd(u_i,v_i)=1$ for some $2r\le i\le n-1$}]\\
	&\ge\gd(u_1,v_1)\gd(u_{2r},v_{2r})+\gd(u_{2r},v_{2r}),
\end{split}
\]
which is easy to verify.

For the case $u_n\neq v_n$:
\[
\begin{split}
	\#T&=\gd(u_1,v_1)\gd(u_2,v_2)+\cdots\\
	&\quad+\gd(u_{2r-2},v_{2r-2})\gd(u_{2r-1},v_{2r-1})\\
	&\quad+\gd(u_{2r},v_{2r})\gd(u_{2r+1},v_{2r+1})+\cdots\\
	&\quad+\gd(u_{n-2},v_{n-2})\gd(u_{n-1},v_{n-1})\\
	&\quad+\gd(u_1,v_1)\gd(u_{2r},v_{2r})\\
	&\le\#\set{1\le i\le 2r-1\mid \gd(u_i,v_i)=1}\\
	&\quad-[\text{$\gd(u_i,v_i)=1$ for some $1\le i\le 2r-1$}]\\
	&\quad+\#\set{2r\le i\le n-1\mid\gd(u_i,v_i)=1}\\
	&\quad-[\text{$\gd(u_i,v_i)=1$ for some $2r\le i\le n-1$}]\\
	&\quad+\gd(u_1,v_1)\gd(u_{2r},v_{2r})\\
	&=\#\set{1\le i\le n\mid \gd(u_i,v_i)=1}\\
	&\quad-[\text{$\gd(u_i,v_i)=1$ for some $1\le i\le 2r-1$}]\\
	&\quad-[\text{$\gd(u_i,v_i)=1$ for some $2r\le i\le n-1$}]\\
	&\quad+\gd(u_1,v_1)\gd(u_{2r},v_{2r})\\
	&\le\#\set{1\le i\le n\mid \gd(u_i,v_i)=1}\\
	&\quad-[\text{$\gd(u_i,v_i)=1$ for some $1\le i\le n$}],\\
\end{split}
\]
where the last inequality follows since
\[
\begin{split}
	&[\text{$\gd(u_i,v_i)=1$ for some $1\le i\le 2r-1$}]\\
	&\quad+[\text{$\gd(u_i,v_i)=1$ for some $2r\le i\le n-1$}]\\
	&\ge\gd(u_1,v_1)\gd(u_{2r},v_{2r})+[\text{$\gd(u_i,v_i)=1$ for some $1\le i\le n$}],
\end{split}
\]
which is easy to verify.

Combining the three cases, we get the required result.
\end{proof}

\section{Examples}

For a given DIM $f$ of length $n$, the square matrix $[D_{ij}]$ of size $n$ whose component $D_{ij}$ is the number of unordered pairs of binary vectors $u,v\in\Z_2^n$ such that $d(u,v)=i$ and $d(f(u),f(v))=j$ is useful to see how much the map $f$ is distance-increasing. Let us call the matrix the \emph{distance expansion table} of the DIM.

Here we exhibit distance expansion tables of DIMs $z_n$ of length $n=5$, $6$, $7$, $8$, $9$, and $11$ defined by our mapping algorithms given in previous sections. The DIMs $z_5$ and $z_9$ whose distance expansion tables are in Figures \ref{fig4} and \ref{fig8} are defined by the mapping algorithm $B$. Note that $z_5$ is the shortest DIM defined by the algorithm $B$. Likewise, DIMs $z_7$ and $z_{11}$ whose distance expansion tables are in Figures \ref{fig5} and \ref{fig9} are defined by the mapping algorithm $C$. The DIM $z_7$ is the shortest DIM defined by the algorithm $C$.

It is easy to see, as claimed in \cite{kwankyu:2005}, that $z_{2r}$ are actually equivalent to $h_{2m}$ from \cite{chang_2003} when $r=m$ is odd or $r=m=2$. So $z_6$ and $h_6$ have the same distance expansion table shown in Figure \ref{fig5}. The distance expansion table of $z_8$ in Figure \ref{fig7} first appeared in \cite{kwankyu:2005}.

After these concrete examples, we conclude that we now have simple mapping algorithms defining DIMs of all length.

\begin{figure}[h]
\begin{center}\small
\[
\left[
\begin{array}{lllll}
 0 & 64 & 16 & 0 & 0 \\
 0 & 0 & 48 & 112 & 0 \\
 0 & 0 & 0 & 64 & 96 \\
 0 & 0 & 0 & 0 & 80 \\
 0 & 0 & 0 & 0 & 16
\end{array}
\right]
\]
\end{center}
\caption{\label{fig4} Distance expansion table of $z_5$}
\end{figure}

\begin{figure}[h]
\begin{center}\small
\[
\left[
\begin{array}{llllll}
 0 & 192 & 0 & 0 & 0 & 0 \\
 0 & 0 & 192 & 288 & 0 & 0 \\
 0 & 0 & 0 & 192 & 384 & 64 \\
 0 & 0 & 0 & 0 & 192 & 288 \\
 0 & 0 & 0 & 0 & 0 & 192 \\
 0 & 0 & 0 & 0 & 0 & 32
\end{array}
\right]
\]
\end{center}
\caption{\label{fig5} Distance expansion table of $z_6$}
\end{figure}

\begin{figure}[h]
\begin{center}\small
\[
\left[
\begin{array}{lllllll}
 0 & 384 & 64 & 0 & 0 & 0 & 0 \\
 0 & 0 & 352 & 832 & 160 & 0 & 0 \\
 0 & 0 & 0 & 320 & 1280 & 576 & 64 \\
 0 & 0 & 0 & 0 & 352 & 1280 & 608 \\
 0 & 0 & 0 & 0 & 0 & 384 & 960 \\
 0 & 0 & 0 & 0 & 0 & 0 & 448 \\
 0 & 0 & 0 & 0 & 0 & 0 & 64
\end{array}
\right]
\]
\end{center}
\caption{\label{fig6} Distance expansion table of $z_7$}
\end{figure}

\begin{figure}[h]
\begin{center}\small
\[
\left[
\begin{array}{llllllll}
 0 & 1024 & 0 & 0 & 0 & 0 & 0 & 0 \\
 0 & 0 & 1024 & 2560 & 0 & 0 & 0 & 0 \\
 0 & 0 & 0 & 1024 & 4096 & 2048 & 0 & 0 \\
 0 & 0 & 0 & 0 & 1024 & 4608 & 3072 & 256 \\
 0 & 0 & 0 & 0 & 0 & 1024 & 4096 & 2048 \\
 0 & 0 & 0 & 0 & 0 & 0 & 1024 & 2560 \\
 0 & 0 & 0 & 0 & 0 & 0 & 0 & 1024 \\
 0 & 0 & 0 & 0 & 0 & 0 & 0 & 128
\end{array}
\right]
\]
\end{center}
\caption{\label{fig7} Distance expansion table of $z_8$}
\end{figure}

\begin{figure*}[hb]
\begin{center}\small
\[
\left[
\begin{array}{lllllllll}
 0 & 2048 & 256 & 0 & 0 & 0 & 0 & 0 & 0 \\
 0 & 0 & 1792 & 6400 & 1024 & 0 & 0 & 0 & 0 \\
 0 & 0 & 0 & 1536 & 10240 & 8704 & 1024 & 0 & 0 \\
 0 & 0 & 0 & 0 & 1280 & 12800 & 13824 & 4352 & 0 \\
 0 & 0 & 0 & 0 & 0 & 1536 & 12544 & 14848 & 3328 \\
 0 & 0 & 0 & 0 & 0 & 0 & 1792 & 11008 & 8704 \\
 0 & 0 & 0 & 0 & 0 & 0 & 0 & 2048 & 7168 \\
 0 & 0 & 0 & 0 & 0 & 0 & 0 & 0 & 2304 \\
 0 & 0 & 0 & 0 & 0 & 0 & 0 & 0 & 256
\end{array}
\right]
\]
\end{center}
\caption{\label{fig8} Distance expansion table of $z_9$}
\end{figure*}

\begin{figure*}[hb]
\begin{center}\small
\[
\left[
\begin{array}{lllllllllll}
 0 & 10240 & 1024 & 0 & 0 & 0 & 0 & 0 & 0 & 0 & 0 \\
 0 & 0 & 9728 & 39936 & 6656 & 0 & 0 & 0 & 0 & 0 & 0 \\
 0 & 0 & 0 & 9216 & 67584 & 78848 & 13312 & 0 & 0 & 0 & 0 \\
 0 & 0 & 0 & 0 & 8704 & 86016 & 158208 & 76800 & 8192 & 0 & 0 \\
 0 & 0 & 0 & 0 & 0 & 8192 & 97280 & 206848 & 137216 & 22528 & 1024 \\
 0 & 0 & 0 & 0 & 0 & 0 & 8704 & 97280 & 214528 & 133120 & 19456 \\
 0 & 0 & 0 & 0 & 0 & 0 & 0 & 9216 & 90112 & 168960 & 69632 \\
 0 & 0 & 0 & 0 & 0 & 0 & 0 & 0 & 9728 & 73728 & 85504 \\
 0 & 0 & 0 & 0 & 0 & 0 & 0 & 0 & 0 & 10240 & 46080 \\
 0 & 0 & 0 & 0 & 0 & 0 & 0 & 0 & 0 & 0 & 11264 \\
 0 & 0 & 0 & 0 & 0 & 0 & 0 & 0 & 0 & 0 & 1024
\end{array}
\right]
\]
\end{center}
\caption{\label{fig9} Distance expansion table of $z_{11}$}
\end{figure*}
\appendices

%\section*{Acknowledgment}
%
%The author thanks the two referees for useful suggestions and comments.

\bibliographystyle{IEEEtran}
%\bibliography{IEEEabrv,dpm}

\begin{thebibliography}{1}
\providecommand{\url}[1]{#1}
\csname url@rmstyle\endcsname
\providecommand{\newblock}{\relax}
\providecommand{\bibinfo}[2]{#2}
\providecommand\BIBentrySTDinterwordspacing{\spaceskip=0pt\relax}
\providecommand\BIBentryALTinterwordstretchfactor{4}
\providecommand\BIBentryALTinterwordspacing{\spaceskip=\fontdimen2\font plus
\BIBentryALTinterwordstretchfactor\fontdimen3\font minus
  \fontdimen4\font\relax}
\providecommand\BIBforeignlanguage[2]{{%
\expandafter\ifx\csname l@#1\endcsname\relax
\typeout{** WARNING: IEEEtran.bst: No hyphenation pattern has been}%
\typeout{** loaded for the language `#1'. Using the pattern for}%
\typeout{** the default language instead.}%
\else
\language=\csname l@#1\endcsname
\fi
#2}}

\bibitem{chang_2003}
J.-C. Chang, R.-J. Chen, T.~{Kl\o ve}, and S.-C. Tsai, ``Distance-preserving
  mappings from binary vectors to permutations,'' \emph{{IEEE} Trans. Inform.
  Theory}, vol.~49, no.~4, pp. 1054--1059, Apr. 2003.

\bibitem{chang:2005}
J.-C. Chang, ``Distance-increasing mappings from binary vectors to
  permutations,'' \emph{{IEEE} Trans. Inform. Theory}, vol.~51, no.~1, pp.
  359--363, Jan. 2005.

\bibitem{kwankyu:2005}
K.~Lee, ``Cyclic constructions of distance-preserving maps,'' to appear
  in~{IEEE} Trans. Inform. Theory.

\end{thebibliography}

%\begin{biographynophoto}{Kwankyu Lee}
%received the B.Sc.~and M.Sc.~degrees in mathematics in 1998 and 2000, respectively, from Sogang
%University, Seoul, Korea.
%
%He is currently a graduate student at the Department of Mathematics, Sogang University. His
%research interests include algebraic coding theory and number theory.
%\end{biographynophoto}

\end{document}